\documentclass[preprint,prd,aps,showpacs,showkeys,nofootinbib]{revtex4}
\usepackage{graphicx}
\usepackage{dcolumn}
\usepackage{bm}
\begin{document}
\title{The order analysis for the two loop corrections to lepton MDM}
\author{Shu-Min Zhao$^{1,2}$\footnote{zhaosm@hbu.edu.cn}, Xing-Xing Dong$^{1,2}$\footnote{dxx$\_$0304@163.com}, Lu-Hao Su$^{1,2}$, Hai-Bin Zhang$^{1,2}$}
\affiliation{$^1$ Department of Physics, Hebei University, Baoding 071002, China}
\affiliation{$^2$ Key Laboratory of High-precision Computation and Application of Quantum Field Theory of Hebei Province, Baoding 071002, China}
\date{\today}

\begin{abstract}
The experimental data of the magnetic dipole moment(MDM) of lepton($e$, $\mu$) is very exact.
The deviation between the experimental data and the standard model prediction maybe come from new physics contribution.
 In the supersymmetric models,
there are very many two loop diagrams contributing to the lepton MDM. In supersymmetric models, we suppose two mass scales
$M_{SH}$ and $M$ with $M_{SH}\gg M$ for supersymmetric particles. Squarks belong to $M_{SH}$ and the other supersymmetric particles belong to $M$.
We analyze the order of the contributions
from the two loop diagrams. The two loop triangle diagrams
corresponding to the two loop self-energy diagram satisfy Ward-identity, and their contributions possess particular factors.
This work can help to distinguish the important two loop diagrams giving corrections to lepton MDM.
\end{abstract}


\maketitle

\section{introduction}
With the detection of the 125 GeV Higgs boson\cite{Higgs125}, the standard model(SM) achieves great success. However, SM has some short comings: 1. SM can not give masses to neutrinos; 2. SM does not have cold dark matter candidate.
The minimal supersymmetric(SUSY) extension of the standard model(MSSM)\cite{MSSM} has attracted physicists' attentions for more than 30 years.
MSSM has also been extended, and the extensions of MSSM\cite{newmodel} have many particles beyond SM. These new particles give corrections to the studied processes.
For the magnetic dipole moment(MDM)\cite{MDM} of lepton especially muon, the two loop corrections from SUSY particles are important.

In the SM, there are several parts giving the contributions to lepton MDM\cite{SMMDM}
\begin{eqnarray}
a_l^{SM}=a_l^{QED}+a_l^{EW}+a_l^{HAD}.
\end{eqnarray}
$a_l^{QED}$ representing the QED contribution is dominant. While,  $a_l^{EW}$ and $a_l^{HAD}$ denote the electroweak and hadronic contributions, respectively. For muon MDM, the deviation between the SM theoretical prediction and experimental data is about 3.7$\sigma$\cite{expthe}
\begin{eqnarray}
\Delta a_\mu= a_\mu^{exp} - a_\mu^{SM} =(27.4\pm 7.3)\times 10^{-10}.
\end{eqnarray}
This deviation $\Delta a_\mu$ may come from the new physics contribution. In SUSY models, there are many SUSY particles that correct
muon MDM through loop diagrams. As discussed in the works\cite{twoloopSUSY1,twoloopSUSY2}, the two loop SUSY diagrams can contribute importantly to muon MDM.

   With the on-shell renormalization scheme, the numerical calculation of two loop electroweak corrections to the muon MDM has been performed in the SM\cite{numtwoloop}. Physicists show great interests on new physics corrections to muon MDM.
   The authors of Ref.\cite{THDtwoloop} study the 2HDM contribution to the muon MDM and present the complete two loop results.
      It is well known that the corrections from SUSY particles are of interest.
   In the work\cite{feng08prd}, the two loop Barr-Zee type diagrams with heavy fermions in the sub-loop are researched. The rainbow diagrams with heavy fermion sub-loop are also deduced analytically\cite{feng08npb}. The contributions to muon MDM from the two loop triangle diagrams generating from the two loop self-energy diagrams b4$[\chi^0; l; \chi^0; \tilde{L}; \tilde{L}]$, b4$[\chi^0; \nu; \chi^\pm; \tilde{L}; \tilde{\nu}]$ and b4$[\chi^\pm; l; \chi^\pm; \tilde{\nu}; \tilde{\nu}]$ can be found in Ref.\cite{feng06prd}. The b4[F1; F2; F3; S1; S2] type diagram is plotted in the FIG.\ref{TLSABC} with F denoting Dirac(Majorana) particles and S representing scalar particles.
 The authors of Ref.\cite{twoloopjx} give detailed results for the two loop corrections to the muon MDM from fermion/sfermion loops in the MSSM.  The two loop diagrams where an additional photon loop is attached
to a SUSY one loop diagram are called as photonic SUSY two loop corrections, whose corrections to muon MDM are evaluated exactly\cite{twoloopphoton}.

   It is well known to all that,  there are a lot of  two loop diagrams\cite{heavy mass} contributing to muon MDM in SUSY models. So,
   to identify the important two loop diagrams is a meaningful thing. We use a method to get the order of the two loop corrections.
   In this work, the number coefficients are not shown, because it is the estimation and the factors $\frac{x_l^{1/2}}{x_M^{1/2}}=\frac{m_l}{M}$ and $\frac{x_l}{x_M}=\frac{m_l^2}{M^2}$ etc are important. After this introduction, we show the used supposition and analysis method in the section II. Section III is devoted to the obtained factors for the corresponding two loop diagrams. In the last section, we discuss the factors and compare their size.

\section{the method}
\begin{figure}[h]
\setlength{\unitlength}{1mm}
\centering
\includegraphics[width=4.2in]{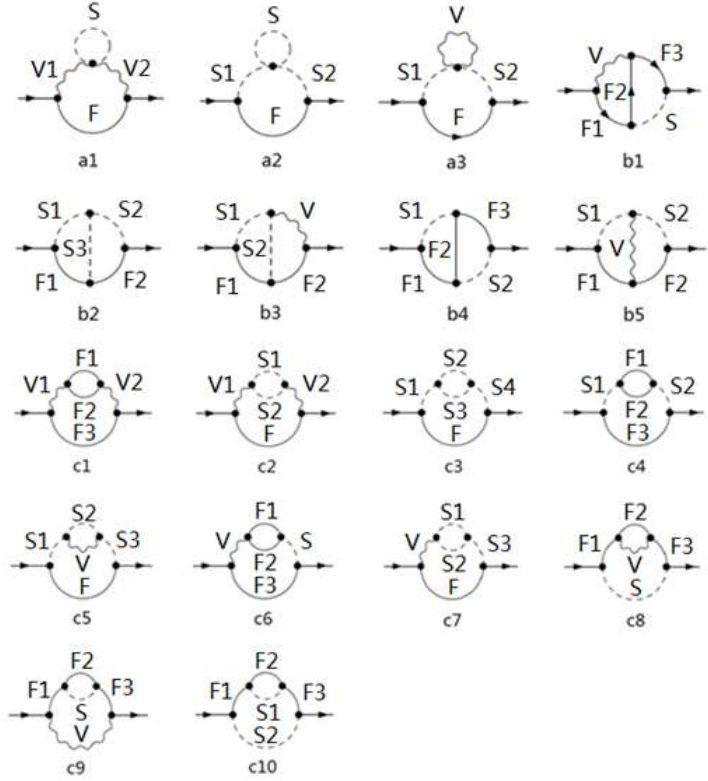}
\vspace{0.0cm}
\caption[]{The two loop self-energy diagrams of topology a, b, c. }\label{TLSABC}
\end{figure}

To make that the analysis is representative, we research the problem in the framework of MSSM.  The two loop triangle diagrams for $\mu\rightarrow \mu +\gamma$ are so many, that we just plot the two loop self-energy diagrams to save space.  The external photon is assumed to be attached to any of the internal, charged particles.
In the FIG.\ref{TLSABC},
the dominant two loop self-energy diagrams for $\mu\rightarrow \mu$ are plotted. For the diagram b1 in the FIG.\ref{TLSABC}, there is another diagram by exchanging the vector boson and scalar boson. However, we do not show it, because its contribution to muon MDM is
Hermitian of the contribution from diagram b1.
Calculating so many two loop diagrams is very difficult and tedious.  So, we use a method to estimate the order of contribution from the two loop diagram. The method is much simpler than the real two loop calculation.
Firstly, we perform the sub-loop integration to get the effective Lagrangian. Secondly, the residual loop integration is performed.

 In the order analysis, the one loop SUSY contributions to lepton MDM are decoupling and possess the factor $\frac{m_l^2}{m^2_{\tilde{l}}}$
with the lepton and slepton masses $m_l, m_{\tilde{l}}$ (some diagrams have an additional enhancement factor $\frac{m_{\tilde{\chi}}}{m_l}$
with some chargino/neutralino mass $m_{\tilde{\chi}}$).
We expect the SUSY corrections can explain the deviation between the measurement and the SM prediction. So, some SUSY particles should be light.
Considering the  current bounds by the LHC, the squark should be heavier ($\gtrsim$ 1.5 TeV). Taking into account the above constraints, we use two mass scales $M$ and $M_{SH}$ for the SUSY particles' masses. $M_{SH}$ represents heavy SUSY particles mass scale, and squarks masses belong to
$M_{SH}$. $M$ is the light SUSY particle mass scale, the masses of the SUSY particles except squarks are supposed as $M$.
$M_{SH}$ is much larger than $M$.
To simplify the calculation and get the factor easily, we adopt another supposition: Considering $m_Z\simeq m_W$\cite{2018pdg}, we use $m_V$ to represent the masses of Z boson and W boson.

  For each two loop self-energy diagram, the corresponding triangle diagrams are obtained by attaching one photon on the internal line and vertex in all possible ways.
The part $\sigma^{\mu\nu}p_\nu$ in the obtained two loop triangle diagrams of the process $\mu\rightarrow \mu+\gamma$ contributes to muon MDM.
The sum of all two loop triangle diagrams belonging to a two loop self-energy diagram satisfies the Ward-identity\cite{wardidentity}.

\begin{figure}[h]
\setlength{\unitlength}{1mm}
\centering
\hspace{1.5cm}\includegraphics[width=3.6in]{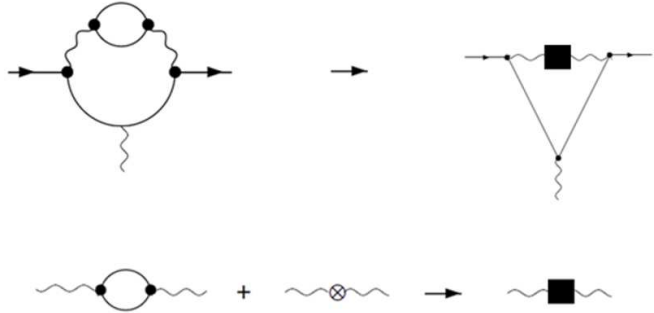}
\vspace{0.0cm}
\caption[]{The schematic diagram for the calculation of the two loop triangle diagram of c1 type.}\label{rainbow}
\end{figure}

To show the method explicitly, we give some examples.  The upper left diagram in FIG.\ref{rainbow} is the two loop triangle diagram of two loop rainbow self-energy diagram c1. It is known that the
contributions from the one loop diagrams are all UV-finite, and at the two loop order only
counterterms for the sub-loops are required. The UV-divergent term comes from the sub-loop that is the one loop self-energy diagram of vector boson. The counter term of the sub-loop is denoted by $\otimes$ in on-shell scheme\cite{onshell}. Adding the counter term, one can obtain the renormalized vector boson self-energy represented by the black square  in the low right diagram of the FIG.\ref{rainbow}. Then we study the effective triangle diagram in the upper right part of the FIG.\ref{rainbow}. It is easy to perform the calculation and obtain the factor of the contribution to muon MDM. Using on-shell scheme, we obtain the decoupling results.

\begin{figure}[h]
\setlength{\unitlength}{1mm}
\centering
\hspace{1.3cm}\includegraphics[width=3.6in]{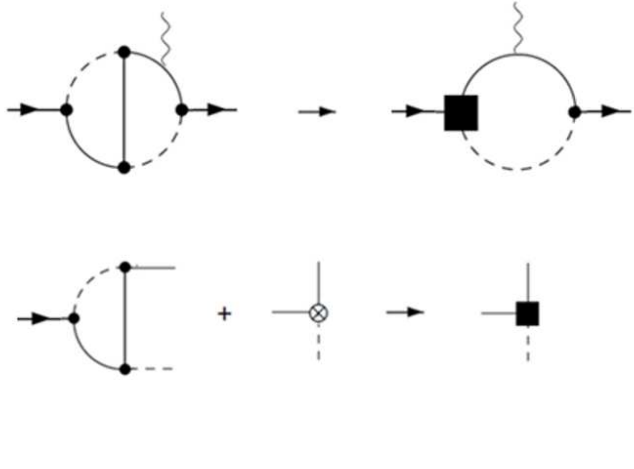};
\vspace{0cm}
\caption[]{The schematic diagram for the calculation of the two loop triangle diagram belonging to b4 type.}\label{TLbing}
\end{figure}

The two loop triangle diagram in the upper left part of the FIG.\ref{TLbing} is UV-divergent which is caused by the sub-loop (one loop
triangle diagram) plotted in the low left part of FIG.\ref{TLbing}.  In the FIG.\ref{TLbing}, the low middle part represented by the $\otimes$ is the counter term of the one loop triangle diagram. The black square in the low right part of the FIG.\ref{TLbing}
denotes the renormalized one loop triangle diagram, which represents the effective vertex. After this operation, the two loop triangle diagram in the upper left part of FIG.\ref{TLbing} is simplified as the effective one loop diagram including black square in the upper right part of FIG. \ref{TLbing}. Then the factor of this diagram's contribution to muon MDM can be obtained.

\section{the analysis}
To show the concrete analysis of the factor, we study the two loop diagrams in the MSSM.
When people insert SUSY particles into the two loop generic diagrams, a lot of MSSM two loop diagrams will be produced.
A two loop self-energy diagram can generate several two loop triangle diagrams.
To denote the two loop diagrams economically, we show our notation for two loop self-energy diagram in the generic form.
For a1[V1; S; V2; F], V represents virtual vector boson, S denotes virtual scalar boson, F is virtual Dirac(Majorana) particle.
The other two loop self-energy diagrams are a2[S1; S; S2; F], a3[S1; S2; V; F], b1[F1; F2; F3; V; S], b2[F1; F2; S1; S2; S3], b3[F1; F2; S1; S2; V],
b4[F1; F2; F3; S1; S2], b5[F1; F2; S1; S2; V], c1[F1; F2; F3; V1; V2], c2[S1; S2; V1; V2; F], c3[S1; S2; S3; S4; F],
c4[F1; F2; F3; S1; S2], c5[S1; S2; S3; V; F], c6[F1; F2; F3; V; S], c7[S1; S2; S3; V; F], c8[F1; F2; F3; V; S], c9[F1; F2; F3; S; V],
c10[F1; F2; F3; S1; S2].

1.
The two loop barr-zee type diagrams $\Big($ c6$[\chi^\pm;~ \chi^\pm;~\mu;~(\gamma,Z);~H^0 ]$ and c6$[\chi^0;~\chi^\pm;~ \nu;~ W^\pm;~H^\pm]\Big)$ give contributions to muon MDM, that are researched in detail by the author Feng\cite{feng08prd}.
Supposing $\chi^\pm$ and $\chi^0$ with the same mass $M$, the analytic results are simplified, then the order factor can be obtained.
The explicit form of the factor is $\frac{x_l}{x_M^{1/2}x_V^{1/2}}$. The notation is $x=\frac{m^2}{\Lambda^2}$ and $\Lambda$ is mass scale. In our method, discarding the vertex coupling we obtain the factor $\frac{x_l^{1/2}}{x_M^{1/2}}$ from loop calculation. In Barr-Zee type diagram, there is $\bar{l}-H-l$ coupling which has the
suppression factor $\frac{m_l}{m_V}=\frac{x_l^{1/2}}{x_V^{1/2}}$ in the vertex. Combining the two factors, we get the order factor $\frac{x_l^{1/2}}{x_M^{1/2}}\times \frac{x_l^{1/2}}{x_V^{1/2}}=\frac{x_l}{x_M^{1/2}x_V^{1/2}}$ that is same as the factor of Feng's result.

2.
The contributions to muon MDM from the two loop diagrams  $\Big($ c1$[\chi^\pm;~ \chi^\pm;~ \mu;~ \gamma;~ (\gamma,~Z)]$, c2$[S;~ S;~ \gamma;~ (\gamma,~Z);~ \mu]$, a1$[\gamma; ~S;~ (\gamma,~Z); ~\mu]\Big)$
 with S denoting the charged scalar particles ($\tilde{L},~ H^\pm$)
 have the order $\frac{x_l}{x_M}$. The rainbow diagrams with heavy fermion in the sub-loop and the two vector bosons being ($\gamma,~\gamma$) and ($\gamma,~Z$)
have been studied in Ref.\cite{feng08npb}, where the factor $\frac{x_l}{x_M}$ in the ($\gamma,~\gamma$) type diagrams is obvious.
Supposing charginos with same mass $M$, the analytic results for two loop rainbow diagrams with two vector bosons$(\gamma,Z)$ in Ref.\cite{feng08npb} are simplified, from which the factor $\frac{x_l}{x_M}$ is extracted. Besides these diagrams,
we find that the diagrams $\Big($c2$[S;~ S;~ \gamma;~ (\gamma,~Z);~ \mu],~$a1$[\gamma; ~S;~ (\gamma,~Z); ~\mu]\Big)$ contribute to muon MDM with the factor $\frac{x_l}{x_M}$.

3.
In Ref.\cite{feng08npb}, the authors research muon MDM corrections from the rainbow diagrams $\Big($c1$[\chi^\pm;~ \chi^\pm;~ \mu;~ Z;~ Z]$ and c1$[\chi^0;~ \chi^\pm;~ \nu;~ W;~ W]\Big)$ that have two heavy vector bosons $ZZ(WW)$ and  heavy fermion sub-loop. From their analytic results, one can get the factor $\frac{x_l}{x_V}$ after the simplification.
In fact, the vector bosons in two loop diagrams $\Big($ c1$[\chi^\pm;~ \chi^\pm;~ \mu;~ Z;~ Z]$, c2$[S;~ S;~ Z;~ Z;~ \mu]$, a1$[Z;~ S; ~Z;~ \mu]$ with $S=\tilde{L},~ H^\pm, ~H^0, ~\tilde{\nu}\Big)$ are just $Z$ bosons. Their sub-diagrams belong to $Z$ self-energy diagrams, that are UV-divergent. To obtain finite and decoupling results, on-shell subduction scheme is used. The sub-diagrams of $\Big($c1$[\chi^0;~ \chi^\pm;~ \nu;~ W;~ W],~ $c2$[S1 ;~ S2 ;~ W;~ W;~ \nu]$ with $(S1,~S2)=(\tilde{\nu},~\tilde{L});~(H^0, ~H^{\pm}), ~$a1$[W;~ S;~ W;~ \nu]$ with $S=\tilde{L}, ~H^\pm, ~H^0, ~\tilde{\nu}\Big)$
 are the self-energy diagrams of W, whose treatment is similar as $Z$ condition. After the estimation, these type two loop diagrams also give corrections to muon MDM with the factor $\frac{x_l}{x_V}$.

4.
This type diagrams$\Big($b2$[\chi^\pm;~ \chi^\pm;~ \tilde{\nu};~\tilde{\nu};~ H^0],  ~$b2$[\chi^0;~ \chi^0;~ \tilde{L};~\tilde{L};~ H^0],~
$b2$[\chi^0;~ \chi^\pm;~ \tilde{L};~\tilde{\nu};~ H^\pm]\Big)$ have the vertex $S-H-S$ possessing mass dimension, which is supposed as $\lambda_{HSS}$.
Considering the Feynman rules in Ref.\cite{MSSM}, the order of $\lambda_{HSS}$ should be in the region $v(250 {\rm GeV})\sim M$.
After the analysis, the factor from loop calculation is  $\frac{x_l^{1/2}}{x_M^{1/2}M}$. Adding the mass dimension parameter $\lambda_{HSS}$ in the vertex,  the total factor of these type diagrams is $\frac{x_l^{1/2}\lambda_{HSS}}{x_M^{1/2}M}$, which is smaller than $\frac{x_l^{1/2}}{x_M^{1/2}}$ because $\frac{\lambda_{HSS}}{M}\leq1$.

5. There are many two loop diagrams contributing to muon MDM with the factor $\frac{x_l^{1/2}}{x_M^{1/2}}$. According to the topology,  these diagrams can be divided into three types: 1. the sub-loop is triangular diagram $\Big($b1$[\mu;~ \chi^0;~ \chi^0;~ Z;~ \tilde{L}]$, b4$[\chi^0; ~\nu;~ \chi^{\pm}; ~\tilde{L};~ \tilde{\nu}]$,  b5$[\chi^{\pm};~ \chi^{\pm};~ \tilde{\nu};~ \tilde{\nu};~ Z]$, b5$[\chi^{0}; ~\chi^{0};~ \tilde{L};~ \tilde{L};~ Z]$, b5$[\chi^{0}; ~\chi^{\pm};~ \tilde{L};~ \tilde{\nu};~ W]\Big)$; 2. the sub-loop is the self-energy diagram of fermion $\Big($c8$[\chi^{\pm};~ \chi^{\pm};~ \chi^{\pm};~ (\gamma,~Z);~ \tilde{\nu}]$, c8$[\chi^{0}; ~\chi^{0};~ \chi^{0}; ~Z; ~\tilde{L}]$, c8$[\chi^{\pm};~ \chi^{0}; ~\chi^{\pm};~ W;~ \tilde{\nu}]$, c8$[\chi^{0}; ~\chi^{\pm};~ \chi^{0};~ W; ~\tilde{L}]$, c9$[\mu;~ \chi^{\pm};~ \mu;~ \tilde{\nu};~ (\gamma,~Z)]$, c9$[\mu;~ \chi^{0}; ~\mu;~ \tilde{L}; ~(\gamma,~Z)]$, c9$[\nu; ~\chi^{\pm};~ \nu; ~\tilde{L}; ~W]$, c9$[\nu;~ \chi^{0};~ \nu; ~\tilde{\nu};~ W]$,  c10$[\chi^\pm; ~\chi^\pm; ~\chi^\pm; ~H^0;~ \tilde{\nu}]$, c10$[\chi^\pm; ~\chi^0; ~\chi^\pm; ~H^\pm;~ \tilde{\nu}]$, c10$[\chi^0; ~\chi^0;~ \chi^0;~ H^0; ~\tilde{L}]$, c10$[\chi^0;~ \chi^\pm; ~\chi^0;~ H^\pm;~ \tilde{L}]\Big)$; 3. the sub-loop is the self-energy diagram of scalar boson $\Big($a2$[\tilde{L}; ~(\tilde{\nu},~\tilde{L},~H^0,~H^\pm);~ \tilde{L};~ \chi^\pm]$, a2$[\tilde{\nu};~ (\tilde{\nu},~\tilde{L},~H^0,~H^\pm);~ \tilde{\nu};~ \chi^0]$, a3$[\tilde{L};~ \tilde{L};~ (\gamma, ~Z,~ W);~ \chi^0]$, a3$[\tilde{\nu};~ \tilde{\nu};~ ( Z,~ W); ~\chi^\pm]$, c4$[\nu; ~\chi^0; ~\chi^{\pm};~ \tilde{\nu};~ \tilde{\nu}]$, c4$[\mu;~ \chi^{\pm}; ~\chi^{\pm};~ \tilde{\nu};~ \tilde{\nu}]$, c4$[\mu; ~\chi^0;~ \chi^{0};~ \tilde{L}; ~\tilde{L}]$, c4$[\nu;~ \chi^{\pm};~ \chi^{0}; ~\tilde{L};~ \tilde{L}]$, c5$[\tilde{L};~ \tilde{L}; ~\tilde{L}; ~(\gamma,~ Z); ~\chi^0]$, c5$[\tilde{\nu}; ~\tilde{\nu};~ \tilde{\nu};~  Z;~ \chi^\pm]$, c5$[\tilde{\nu}; ~\tilde{L}; ~\tilde{\nu}; ~ W; ~\chi^\pm]$, c5$[\tilde{L};~ \tilde{\nu};~ \tilde{L}; ~ W; ~\chi^0]\Big)$. Here, the studied SUSY particles possess same mass $M$, and
the analysis of these three type diagrams are not difficult. At first, we perform the loop integration of the sub-loop and obtain the effective
coefficients and operators. Secondly, we calculate the remaining loop integration. At last, the factor of the contributions to muon MDM is extracted.  In the end, we find that their contributions all have the factor $\frac{x_l^{1/2}}{x_M^{1/2}}$.

6.
The sub-diagrams of this type two loop self-energy diagrams $\Big($c$10[\chi^0;~ F;~ \chi^0;~ \tilde{S};~ \tilde{L}]$ with $(F,~\tilde{S})=(\nu,~\tilde{\nu}),~ (l,~\tilde{L})$ and
c$10[\chi^\pm;~ F; ~\chi^\pm;~ \tilde{S};~ \tilde{\nu}]$ with $(F,~\tilde{S})=(\nu,~\tilde{L}),~ (l,~\tilde{\nu})\Big)$ are the fermion self-energy diagrams with virtual scalar particles.
These diagrams give corrections to muon MDM with the factor $\frac{x_l^{1/2}}{x_M^{1/2}}\frac{m_F}{M}$. Here,
$m_F$ represents the mass of virtual fermion including SM fermion, $\chi^{\pm}$ and $\chi^0$ in the two loop diagram.  Obviously, $\frac{m_F}{M}\leq1$ and  the factor $\frac{x_l^{1/2}}{x_M^{1/2}}\frac{m_F}{M}\leq \frac{x_l^{1/2}}{x_M^{1/2}}$.

7.
The mass of lepton is much smaller than the $m_V$. That is to say $\frac{x_l}{x_V}\ll1$, and we
can expand the results corresponding to $\frac{x_l^{1/2}}{x_M^{1/2}}$ and $\frac{x_l}{x_V}$. For the
contributions from the diagrams $\Big($b1$[\mu;~ \chi^\pm;~ \chi^\pm; ~(\gamma,~ Z);~ \tilde{\nu}]$,
b1$[\nu;~ \chi^0;~ \chi^\pm;~ W; ~\tilde{\nu}]$, b1$[\nu;~ \chi^\pm;~ \chi^0;~ W;~ \tilde{L}]$,
b3$[\chi^\pm;~ \mu;~ \tilde{\nu};~ \tilde{\nu}; ~Z]$, b3$[\chi^0;~ \mu; ~\tilde{L}; ~\tilde{L}; ~(\gamma,~ Z)]$,
b3$[\chi^0;~ \nu; ~\tilde{L};~ \tilde{\nu}; ~W]$, b3$[\chi^\pm; ~\nu;~ \tilde{\nu}; ~\tilde{L};~ W]\Big)$, some parts have the factor $\frac{x_l^{1/2}}{x_M^{1/2}}$ and the other parts
have the factor $\frac{x_l}{x_V}$ after our expansion.

8. This type diagrams $\Big($b4$[\mu;~ \chi^\pm;~ \chi^\pm;~ H^0; ~\tilde{\nu}]$, b4$[\mu; ~\chi^0; ~\chi^0;~ H^0; ~\tilde{L}]$,
b4$[\nu; ~\chi^0;~ \chi^\pm; ~H^\pm; ~\tilde{\nu}]$, b4$[\nu; ~\chi^\pm; ~\chi^0;~ H^\pm;~ \tilde{L}]\Big)$ are very similar as the diagrams $\Big($  b1$[\mu;~ \chi^\pm;~ \chi^\pm; ~(\gamma,~ Z);~ \tilde{\nu}]$,
 b1$[\nu;~ \chi^0;~ \chi^\pm;~ W; ~\tilde{\nu}]$, b1$[\nu;~ \chi^\pm;~ \chi^0;~ W;~ \tilde{L}]
 \Big)$. If the vector bosons $\gamma,Z,W$ are replaced by $H^0,H^{\pm}$ respectively, we can obtain the b4 type diagrams.
Without considering the vertex couplings, we obtain the factors $\frac{x_l^{1/2}}{x_M^{1/2}}$ and $\frac{x_l}{x_V}$ of the $b1$ type diagrams contributions. This condition is almost same as the condition of the diagrams $\Big($b1$[\mu;~ \chi^\pm;~ \chi^\pm; ~(\gamma,~ Z);~ \tilde{\nu}]\dots$ and
b3$[\chi^\pm;~ \mu;~ \tilde{\nu};~ \tilde{\nu}; ~Z]\dots\Big)$ with the factors $\frac{x_l^{1/2}}{x_M^{1/2}}$ and $\frac{x_l}{x_V}$ in the up section.
At last, we should consider the suppression factor $\frac{x_l^{1/2}}{x_V^{1/2}}$ in the vertex $\bar{l}-H-l$. Therefore, the final order factors are $\frac{x_l}{x_M^{1/2}x_V^{1/2}}$ and $\frac{x_l^{3/2}}{x_Hx_V^{1/2}}$. Here $x_H=\frac{m_H^2}{\Lambda^2}$, and $m_H$ is heavy Higgs mass.

9. The Higgs (including the 125GeV Higgs boson and goldstone) are not of the same masses. In this work, we distinguish the goldstone boson, light CP-even Higgs and Heavy Higgs.  $m_{h}$ denotes light CP-even Higgs with mass ($m_h$=125.1GeV), as well as $m_{H}$ represents heavy Higgs mass.
For the diagrams $\Big($c7$[H^0;~ H^0; ~H^0;~ (\gamma, ~Z); ~\mu]$,
c7$[\tilde{\nu}; ~\tilde{\nu};~ H^0; ~Z; ~\mu]$, c7$[S; ~S;~ H^0; ~(\gamma,~ Z); ~\mu]$ with $S=\tilde{L}, ~H^{\pm}$ and c7$[S1; ~S2;~ H^\pm; ~W;~ \nu]$ with $(S1, ~S2)=
(\tilde{\nu},~ \tilde{L});~ (H^\pm,~ H^0)\Big)$, we perform the loop integration and expand the contributions to muon MDM. The results can be departed into two parts: one part has the factor $\frac{x_l^{1/2}}{x_M^{1/2}M}$ and the other part has the factor $\frac{x_l^{1/2}}{x_H^{1/2}m_H}$. When the Higgs are goldstone and light CP-even Higgs, the extracted factor is $\frac{x_l^{1/2}}{x_M^{1/2}M}$. If the two loop diagrams include heavy Higgs with $m_H\gtrsim M$, the factor is $\frac{x_l^{1/2}}{x_H^{1/2}m_H}$.
In these type diagrams, the two vertex couplings must be taken into account: $S-H-S$ vertex coupling with mass dimension $\lambda_{HSS}$ and $\bar{l}-H-l$ vertex coupling with the supression factor $\frac{m_l}{m_V}$.
In the end, these type diagrams give muon MDM corrections with two factors: $\frac{x_l\lambda_{HSS}}{x_V^{1/2}x_M^{1/2}M}$ and $\frac{x_l\lambda_{HSS}}{x_V^{1/2}x_H^{1/2}m_H}$.  In rough estimation, $\frac{\lambda_{HSS}}{m_H}\lesssim 1$.

10. For the diagrams  $\Big($c3$[\tilde{\nu};~ H^0; ~\tilde{\nu};~ \tilde{\nu};~ \chi^{\pm}]$,  c3$[\tilde{\nu};~ H^\pm; ~\tilde{L};~ \tilde{\nu}; ~\chi^{\pm}]$, c3$[\tilde{L}; ~H^0;~ \tilde{L};~ \tilde{L}; ~\chi^{0}]$, c3$[\tilde{L}; ~H^\pm; ~\tilde{\nu}; ~\tilde{L}; ~\chi^{0}]\Big)$, we obtain two factors from the loop integration, which include $\frac{x_l^{1/2}}{x_M^{1/2}M^2}$. The other factor has relation with the Higgs masses. If the Higgs are the goldstone boson and light CP-even Higgs(125GeV), the factor $\frac{x_l^{1/2}}{x_M^{1/2}M^2}$ is same as the front factor. If the Higgs masses are heavier than $M$, the corresponding factor is$\frac{x_l^{1/2}}{x_M^{1/2}m_{H}^2}$.  There are two $S-H-S$ vertexes with mass dimension parameter $\lambda_{HSS}$ in these two loop diagrams.
Taking into account $\lambda_{HSS}$, we gain the final factors  $\frac{x_l^{1/2}\lambda_{HSS}^2}{x_M^{1/2}M^2}$ and $\frac{x_l^{1/2}\lambda_{HSS}^2}{x_M^{1/2}m^2_H}$ of the contributions to muon MDM.

11. As discussed in Ref.\cite{log}, the c10[F1; F2; F3; S1; S2] type diagram with large sfermion masses can give non-decoupling and logarithmically enhanced corrections to muon MDM. In our supposition, squarks are very heavy, and their masses belong to $M_{SH}$ with the
relation $M_{SH}\gg M$. The two loop self-energy diagrams $\Big($c$10[\chi^0;~ F;~ \chi^0;~ \tilde{S};~ \tilde{L}]$ with $(F,~\tilde{S})= (u,~\tilde{U}) ,~(d,~\tilde{D})$ and
c$10[\chi^\pm;~ F; ~\chi^\pm;~ \tilde{S};~ \tilde{\nu}]$ with $(F,~\tilde{S})=(u,~\tilde{D}) ,~(d,~\tilde{U})\Big)$ possess heavy squarks, and the factor of their contributions should include the large logarithm function. After derivation, we obtain the factor  $\frac{x_l^{1/2}}{x_M^{1/2}}\frac{m_F}{M}\log{x_{SH}}$. $m_F$ represents the mass of virtual fermion and $\frac{m_F}{M}$ is not larger than 1. The logarithmically enhanced factor is $\log{x_{SH}}$, with $x_{SH}=\frac{M^2_{SH}}{\Lambda^2}$.

\section{discussion and conclusion}

 In this work, for supersymmetric particles we suppose two mass scales
$M_{SH}$ and $M$ with the relation $M_{SH}\gg M$.
  Taking into account of the vertex coupling, the orders for the two loop diagrams contributing to lepton MDM are analyzed here. Their contributions have particular factors representing the order, and these factors are collected here. The Barr-Zee type factor is $\frac{x_l}{x_M^{1/2}x_V^{1/2}}$. The other type factors are $\frac{x_l^{1/2}}{x_M^{1/2}}$, $\frac{x_l^{1/2}\lambda_{HSS}}{x_M^{1/2}M}$, $\frac{x_l^{1/2}\lambda_{HSS}^2}{x_M^{1/2}M^2}$, $\frac{x_l^{1/2}\lambda_{HSS}^2}{x_M^{1/2}m^2_H}$, $\frac{x_l^{1/2}}{x_M^{1/2}}\frac{m_F}{M}$,  $\frac{x_l}{x_V}$, $\frac{x_l\lambda_{HSS}}{x_V^{1/2}x_M^{1/2}M}$, $\frac{x_l\lambda_{HSS}}{x_V^{1/2}x_H^{1/2}m_H}$, $\frac{x_l}{x_M}$, $\frac{x_l^{3/2}}{x_Hx_V^{1/2}}$ and $\frac{x_l^{1/2}}{x_M^{1/2}}\frac{m_F}{M}\log{x_{SH}}$. These factors except $\frac{x_l}{x_V}$ and $\frac{x_l^{1/2}}{x_M^{1/2}}\frac{m_F}{M}\log{x_{SH}}$ become small with the enlarging masses of SUSY
particles and Higgs.

In Ref.\cite{twoloopSUSY1}, the authors have reached the two loop diagrams including barr-zee type diagrams with Fermion sub-loop and two loop rainbow diagrams with Fermion sub-loop. These two loop diagrams have been researched in great detail in Refs.\cite{feng08prd,feng08npb} and the analytic results are obtained. Our order analysis is same with the results of Refs.\cite{feng08prd,feng08npb}.  For the two loop barr-zee type diagrams with scalar sub-loop in Ref.\cite{twoloopSUSY1}, their factor is $\frac{x_l}{x_M^{1/2}x_H^{1/2}}$ with $m_H$ representing heavy Higgs mass. In our analysis, the two loop barr-zee type diagrams with scalar sub-loop belong to c7 type, whose factors are  $\frac{x_l\lambda_{HSS}}{x_V^{1/2}x_M^{1/2}M}$ and $\frac{x_l\lambda_{HSS}}{x_V^{1/2}x_H^{1/2}m_H}$. When the Higgs is heavy, our factor is $\frac{x_l\lambda_{HSS}}{x_V^{1/2}x_H^{1/2}m_H}$ which is almost same as the factor $\frac{x_l}{x_M^{1/2}x_H^{1/2}}$ in Ref.\cite{twoloopSUSY1}.

  In Ref.\cite{twoloopjx} the two loop diagrams of c10 type with scalar leptons are calculated, where the factor is gained as $\frac{x_l}{x_M}$. Our factor from slepton contribution in the c10 type diagrams is $\frac{x_l^{1/2}}{x_M^{1/2}}\frac{m_F}{M}$. Corresponding to the two loop diagrams
in Ref.\cite{twoloopjx}, $m_F=m_l$, therefore our factor  $\frac{x_l^{1/2}}{x_M^{1/2}}\frac{m_l}{M}=\frac{x_l}{x_M}$ is same as the factor in Ref.\cite{twoloopjx}.
The corrections from c10 type diagram with large squarks are non-decoupling and logarithmically enhanced. In this condition, we show the factor with large logarithm function as $\frac{x_l^{1/2}}{x_M^{1/2}}\frac{m_F}{M}\log{x_{SH}}$. This factor is consistent with the results
of Refs.\cite{twoloopjx,log} for the heavy scalar particles.
The studied two loop diagrams in Ref.\cite{twoloopphoton} belong to the types b1, b3, c8 and c9. From their results, the obtained factor is $\frac{x_l^{1/2}}{x_M^{1/2}}$, which is completely same as the factor in our analysis.

The small factors are $\frac{x_l}{x_M}$ and $\frac{x_l^{3/2}}{x_Hx_V^{1/2}}$. If the two loop contributions possess just the
both factors, their contributions can be neglected safely.  $\frac{x_l^{1/2}}{x_M^{1/2}}$ and $\frac{x_l}{x_V}$ are the large factors.
 Because $\frac{\lambda_{HSS}}{M}$, $\frac{\lambda_{HSS}^2}{M^2}$, $\frac{\lambda_{HSS}^2}{m^2_H}$ and $\frac{m_F}{M}$ are not more than 1, the factors ($\frac{x_l^{1/2}\lambda_{HSS}}{x_M^{1/2}M}$,$\frac{x_l^{1/2}\lambda_{HSS}^2}{x_M^{1/2}M^2}$, $\frac{x_l^{1/2}\lambda_{HSS}^2}{x_M^{1/2}m^2_H}$,$\frac{x_l^{1/2}}{x_M^{1/2}}\frac{m_F}{M}$) should not be bigger than the factor $\frac{x_l^{1/2}}{x_M^{1/2}}$.
The factor $\frac{x_l}{x_M^{1/2}x_V^{1/2}}$ for the famous barr-zee type two loop diagrams is of the middle order, which is similar as the factors $\frac{x_l\lambda_{HSS}}{x_V^{1/2}x_M^{1/2}M}$ and $\frac{x_l\lambda_{HSS}}{x_V^{1/2}x_H^{1/2}m_H}$. The non-decoupling factor $\frac{x_l^{1/2}}{x_M^{1/2}}\frac{m_F}{M}\log{x_{SH}}$ is special.
This work is the order analysis for the two loop diagrams. That is to say the values for the parameters rely on the concrete models.
It is in favor of reseaching two loop corrections to the lepton MDM in the models beyond SM.

\begin{acknowledgments}
This work is supported by National Natural Science Foundation of China (NNSFC) (No. 11535002, No. 11705045), Natural Science Foundation of Hebei Province
(A2020201002)
Post-graduate's Innovation Fund Project of Hebei Province
 (No. CXZZBS2019027),  and the youth top-notch talent support program of the Hebei Province.
\end{acknowledgments}

 \end{document}